\documentclass{article}
\usepackage{spconf, enumitem}
\usepackage{amsmath}
\usepackage{amsfonts}
\usepackage{tikz}
\usetikzlibrary{matrix,shapes,arrows,positioning,chains, calc}

\title{Privacy preserving distance computation using somewhat-trusted third parties}
\name{Abelino Jimenez  and  Bhiksha Raj}
\address{Carnegie Mellon University, Pittsburgh, PA, USA \\
                             {\{abjimenez,bhiksha\}@cmu.edu}}

\begin{document}
\ninept
                 \maketitle
                 \begin{abstract}
A critically important component of most signal processing procedures is that of computing the distance between signals. In multi-party processing applications where these signals belong to different parties, this introduces privacy challenges. The signals may themselves be private, and the parties to the computation may not be willing to expose them. Solutions proposed to the problem in the literature generally invoke homomorphic encryption schemes, secure multi-party computation, or other cryptographic methods which introduce significant computational complexity into the proceedings, often to the point of making more complex computations requiring repeated computations unfeasible. Other solutions invoke  third parties, making unrealistic assumptions about their trustworthiness. 

In this paper we propose an alternate approach, also based on third party computation, but without assuming as much trust in the third party. Individual participants to the computation ``secure'' their data through a proposed secure hashing scheme with shared keys, prior to sharing it with the third party. The hashing ensures that the third party cannot recover any information about the individual signals or their statistics, either from analysis of individual computations or their long-term aggregate patterns. We provide theoretical proof of these properties and empirical demonstration of the feasibility of the computation.
                 \end{abstract}

                         .
\section{Introduction}
A key signal processing operation is the computation of the distance between two signals. This simple operation can suddenly become challenging if the two signals belong to two mistrustful parties who are unwilling to share them. We must now facilitate the continuation of the operations, without exposing their signals to one another.

More formally stated, two parties, Alice and Bob, have two real valued vectors ${\bf x}_1$ and ${\bf x}_2 \in \mathbb{R}^N$ respectively. The signal processing operation requires the computation of $\|{\bf x}_1 - {\bf x}_2\|$, which may be required by one or both parties. However, at the same time, it is required that none besides Alice learns about ${\bf x}_1$, and similarly none besides Bob is exposed to ${\bf x}_2$.

A number of methods have been suggested in the literature to solve this problem. These have generally involved the use of secure multi-party computation protocols [GCS] \cite{manaspathak:smc} or fully homomorphic \cite{fhm:latest, RANE:LSHinsecure} which enables the computation of distances over encrypted data. Although this provides the participants with the desired privacy, they are computationally unattractive. They can increase both the communication and the computational cost of computing the distance by several orders of magnitude.

An alternate approach uses {\em trusted third parties} to facilitate the computation. Here,  Alice and Bob interact with a third party, Charlie to compute the distances between their signals with information-theoretic privacy. No party gets additional information despite their computational power. 
%To illustrate this sort of protocols, consider that Alice and Bob have binary vector ${\bf x}_1$ and ${\bf x}_2 \in \{0,1\}^N$ respectively, and they wish to compute the hamming distance between them [REF]. Alice can generate a random vector ${\bf y} \in \{0,1\}^N$ and a permutation $\mathcal{P}$. Then, Alice send $({\bf y}, \mathcal{P})$ to Bob secretly, and both send $\mathcal{P}({\bf x}_i \oplus {\bf y})$ to Charles. Now, Charles can compute the Hamming distance $d$ between the vectors received, which corresponds to the hamming distance between ${\bf x}_1$ and ${\bf x}_2$, and send $d$ to Alice and Bob. Hence, no party can get nothing more than the Hamming distance.
Under this approach, protocols based on secret sharing using polynomials on finite fields have been proposed to compute the euclidean distance \cite{PolyDistance} between privately held real-valued vectors. While secure,  the precision of the computation depends on how well real numbers are represented in the selected field. Moreover, the communication complexity of this kind of protocol is proportional to the dimensionality of the data, which is problematic when Alice and Bob need to compute a distance between high dimensional vectors in presence of communication constraints.

{\em Even} if the above concerns were somehow allayed, an additional problem arises from the fundamental definition of the basic problem itself. The outcome of the computation -- the distance -- itself reveals information about the signals. The party who obtains this result, which may be either Alice or Bob, or both of them, learns about the relationship  between their signals.  In trusted third-party settings, Charlie may get to know this distance too. Ideally, while this ``leakage'' of information cannot be avoided, we would like to limit it. In particular, it is preferable that the third party Charlie not be exposed to the information at all. 

In this paper we propose several third-party based protocols for the computation of distances between signals, which address the above issues. We propose to encode the signals through a {\em modular} hashing scheme \cite{Jimenez:wifs} with cryptographic properties that ensure that none of the parties can recover the original signal from the hashes, without the appropriate keys. The hashes have the property that they permit {\em limited} release of the information between vectors -- the true distance is only revealed if the vectors are sufficiently close and the distance between them lies below a threshold, but is naturally effectively thresholded when the distance is greater, ensuring that even the parties that {\em do} get the distance only get as much as is required for effective computation of the signal processing processes, but not enough to chart out the other person's data. We impose protocols on top of the hashing which ensure that the third party cannot gather any information besides the limited distance computed. When the third party may be trusted to know this limited distance then, unlike other cryptographic or secret sharing based schemes, the communication overhead may be made independent of the dimension of the signals, permitting control of the precision of the distance estimate. Finally, we also analyze how the protocols may be modified in order to not expose even the limited distance to the third party. 

In the next section (Section 2) we discuss the hashing scheme. In Section 3 we describe our protocols and in Section 4 we explain how they may be extended to hide the distance from the third party (Charlie). Finally we present our conclusions in Section 5.

\section{Limited Distance Computation through Modular Hashes}
Our proposed approach requires Alice and Bob to transform their signals into hashes prior to further processing. We now describe the transform. This transformation may be considered as a \textit{modular} variant of the $p$-stable Locality Sensitive Hashing (LSH) \cite{LSH:Indyk} \cite{LSHfordistance}. However, unlike the standard LSH, the introduction of modularity helps to generate {\em uninformative hashes}, i.e. the distribution of the resulting hash does not depend on its input. Furthermore, the euclidean distance between two signals may be estimated through comparison of their hashes, provided the distance is shorter than some threshold. 

\subsection{Modular Hashing}
\noindent{\bf Definition:} Let $k$ be an integer larger than 1, $A$ be an $M \times N$ matrix,  $U \in [0,k]^M$, and $\mathbb{Z}_k$ be the set of integers $\{0, 1, \ldots, k-1\}$. We define a {\em Modular Hash} as the function $Q_{k,A,U} \, : \, \mathbb{R}^N \rightarrow \mathbb{Z}_k^M$ as
\begin{equation}
Q_{k,A,U}({\bf x}) = \left\lfloor A {\bf x}+ U \right\rfloor (\textrm{mod} \, \, k)
\label{eq:smh}
\end{equation}
where the floor function and modulo are component-wise.

\vspace{0.1cm}
\noindent{\bf Definition:} If $A$ and $U$ are randomly selected, we say that $Q_{k,A,U}$ is a \textit{Secure Modular Hash} (SMH) if 
\begin{enumerate}
\item $\forall {\bf x} \in \mathbb{R}^N$, $Q_{k,A,U}({\bf x})_i$ is independent of $Q_{k,A,U}({\bf x})_j$ for every $i\neq j$,
\item $\forall i \in\{1,...,M\}, \forall z \in \mathbb{Z}_k$,   $\mathbb{P}\left( Q_{k,A,U}({\bf x})_i = z\right) = \frac{1}{k}$
\end{enumerate}

\noindent One example of SMH is given in \cite{Jimenez:wifs} as follows:

\vspace{0.1cm}
\noindent \textbf{Theorem 1.} $Q_{k,A,U}$ is a SMH if $A$ is randomly generated where its components are i.i.d with
%$$a_{ij} \sim \mathcal{N}\left( 0,1/\delta^2 \right)$$
$$a_{ij} \sim \mathcal{N}\left( 0,\delta^{-2} \right)$$
for some $\delta$, and $U$ is independent of $A$, with i.i.d components
$$u_i \sim \text{Unif}(0,k)$$
 It is  shown in \cite{Jimenez:wifs} that for SMH defined as above, the expected value of the Hamming distance between $Q_{k,A,U}({\bf x}_1)$ and $Q_{k,A,U}({\bf x}_2)$ relates to the actual value of $\|{\bf x}_1 - {\bf x}_2\|$, until the former achieves the value of $1-\frac{1}{k}$, beyond which the hash is information theoretically secure and reveals no information about $\|{\bf x}_1 - {\bf x}_2\|$. We will henceforth assume that we work with this kind of SMH.

\subsection{Limited Euclidean Distance from Secure Modular Hashes}
%Considering $(k,A,U)$ as in Theorem 1, in \cite{Jimenez:wifs} is proved that the expected value of the Hamming distance between $Q_{k,A,U}({\bf x}_1)$ and $Q_{k,A,U}({\bf x}_2)$ depends on the actual value of $\|{\bf x}_1 - {\bf x}_2\|$.

%However, if we need to estimate the value of the euclidean distance, the process may be confusing and new approximation may be required, increasing the error estimation.
The {\em Lee} distance \cite{LeeDistance} between two integers $a,b \in \mathbb{Z}_k$ is the length of the shortest path from $a$ to $b$ along a ring of circumference $k$.
%
%In order to get a more tractable way to estimate the euclidean distance, we are going to take the advantage of the topology of $\mathbb{Z}_k$. In particular, we will work with the Lee distance , this is, if $a$ and $b$ belong to $\mathbb{Z}_k$, the Lee distance between $a$ and $b$ is defined as
$$d_{Lee}(a,b) = \min \left( |a-b|, k - |a-b| \right)$$
%A good illustration for understanding this distance is consider $\mathbb{Z}_k$ as a clock with $k$ hours. Then, the modular distance between two points is the length of the shortest path from one point to the other over the clock.
Using this distance metric, we can obtain the following result:

\vspace{0.1cm}
\noindent \textbf{Theorem 2.} $\forall {\bf x}_1 \, , {\bf x}_2 \in \mathbb{R}^N$, $\forall i \in \{1,...,M\}$,  $\forall k$ even, then we obtain the following relation between the $i^{\rm th}$ components of $Q_{k,A,U}({\bf x}_1)$ and $Q_{k,A,U}({\bf x}_2)$:
\begin{multline*}
 \mathbb{E} \left[ d_{Lee}\left(Q_{k,A,U}({\bf x}_1)_i,Q_{k,A,U}({\bf x}_2)_i \right) \right] = \\
 \frac{k}{4} -  \frac{2k}{\pi^2} \sum\limits_{j=1}^\infty \frac{1}{(2j-1)^2} e^{-2 \left( \frac{\pi \|{\bf x}_1-{\bf x}_2\|(2j-1)}{\delta k}\right)^2}  
\end{multline*}
%
%The following inequality let us understand the functional relation between the actual euclidean distance between ${\bf x}_1$ and ${\bf x}_2$ and the Lee distance between their corresponding hashes
%
%\noindent \textbf{Proposition 1.} 
%\begin{multline*}
%\frac{k}{4}-\frac{k}{4}e^{-2 \left( \frac{\pi \|{\bf x}_1-{\bf x}_2\|}{\delta k}\right)^2} \leq \\
%\mathbb{E}[d_{Lee}\left(Q_{k,A,U}({\bf x}_1)_i,Q_{k,A,U}({\bf x}_2)_i \right)] \leq \\ \frac{k}{4}-\frac{2k}{\pi^2}e^{-2 \left( \frac{\pi \|{\bf x}_1-{\bf %x}_2\|}{\delta k}\right)^2}
%\end{multline*}
Moreover, bounding this expression it is possible to prove that when
$\|{\bf x}_1 - {\bf x}_2 \| \rightarrow \infty$ the described series converges to $\frac{k}{4}$. 

%Nevertheless, this function seems to be complex, and, even though there is a functional relation between $\|{\bf x}_1 - {\bf x}_2\|$ and the expected value of Lee distance between the corresponding hashes, it is not clear how to estimate the actual euclidean distance after applying this kind of technique.

The {\em expected} value of the Lee distance between $Q_{k,A,U}({\bf x}_1)_i$ and $Q_{k,A,U}({\bf x}_2)_i$ approximates the Euclidean distance between ${\bf x}_1$ and ${\bf x}_2$. The following result bounds the error of the approximation.

\vspace{0.1cm}
\noindent \textbf{Proposition 1.}
%\begin{multline*}
%\left| \mathbb{E}[d_{Lee}\left(Q_{k,A,U}({\bf x}_1)_i,Q_{k,A,U}({\bf x}_2)_i \right)] - \sqrt{\frac{2}{\pi \delta^2}} \| {\bf x}_1 - {\bf x}_2 \| \right| \\
%\leq \sqrt{\frac{2}{\pi \delta^2}} \| {\bf x}_1 - {\bf x}_2 \| \exp \left( - \frac{k^2 \delta^2}{8 \|{\bf x}_1 - {\bf x}_2 \|^2} \right)
%\end{multline*}
Let $\delta = \sqrt{\frac{2}{\pi}}$. We define the error 

\vspace{-0.6cm}
\begin{multline*}
 \varepsilon(\|{\bf x}_1 - {\bf x}_2 \|,k) : = \\
 \left| \mathbb{E}[d_{Lee}\left(Q_{k,A,U}({\bf x}_1)_i,Q_{k,A,U}({\bf x}_2)_i \right)] -  \| {\bf x}_1 - {\bf x}_2 \| \right| 
\end{multline*}
The following relation holds:

\vspace{-0.3cm}
\begin{equation}
\varepsilon(\|{\bf x}_1 - {\bf x}_2 \|, k) \leq F(\|{\bf x}_1 - {\bf x}_2 \|, k) 
\end{equation}
where
$$F(t,k) = t \cdot \exp \left( - \frac{k^2}{4 \pi t^2} \right)$$

It is easy to see that $F$ is increasing in $t$ and decreasing in $k$. Even more, for fixed $t$, when $k$ tends to infinity $F(t,k)$ tends to 0. Therefore, we can prove the following proposition:

\vspace{0.3cm}
\noindent \textbf{Proposition 2.} $\forall \epsilon >0$ , $\forall T>0$, $\exists k$ even,  $\forall \|{\bf x}_1 - {\bf x}_2\| <T$
$$ \varepsilon(\|{\bf x}_1 - {\bf x}_2\|,k) < \epsilon$$

\begin{figure}[t]
\centering
\includegraphics[trim={4.5cm 7.5cm 2cm 8cm},clip, scale=0.4]{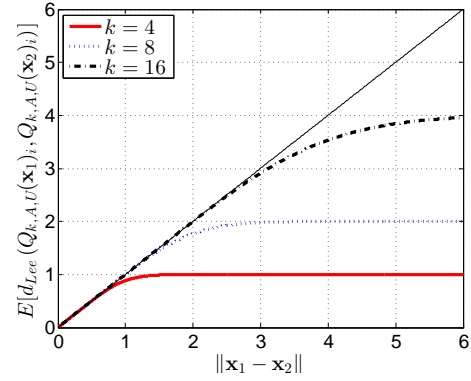}
\caption{Expected value of the Lee Distance between $Q_{k,A,U}({\bf x}_1)_i$ and $Q_{k,A,U}({\bf x}_2)_i$  as a function of $\|{\bf x}_1- {\bf x}_2\|$ using expression given by Theorem 2 and considering $\delta = \sqrt{\frac{2}{\pi}}$.}
\label{fig:TheoreticalCurve}
\vspace{-0.5cm}
\end{figure}

This proposition says that given a threshold $T$ we can find a value of $k$ large enough, such that the difference between  $\|{\bf x}_1 - {\bf x}_2\|$ and the expected value of the Lee distance between the corresponding hashes is as small as we want for all $\|{\bf x}_1 - {\bf x}_2\| < T$. 
%This can be seen in the Figure \ref{fig:bound}, where $F(t,k)$ is plotted for different values of $k$. We can see than while $k$ increases, the threshold of preservation increases for a given level of precision. 
This is illustrated by Figure \ref{fig:TheoreticalCurve} which shows the expression given by Theorem 2 against $\|{\bf x}_1 - {\bf x}_2\|$. The relation is seen to be an identity for distances smaller than a threshold, and then converging to $\frac{k}{4}$. Hence, we can compute an accurate estimate of the euclidean distance between ${\bf x}_1$ and ${\bf x}_2$ directly from the Lee distance between their hashes.

%\begin{figure}[t]
%\centering
%\includegraphics[trim={3cm 8.5cm 3cm 9cm},clip, scale=0.55]{figs/errorConservation.pdf}
%\caption{Upper bound $F(\|{\bf x}_1 - {\bf x}_2\|,k)$ for the difference between $\|{\bf x}_1 - {\bf x}_2\|$ and the expected value of the Lee distance between $Q_{k,A,U}({\bf x}_1)_i$ and $Q_{k,A,U}({\bf x}_2)_i$, as a function of $\|{\bf x}_1 - {\bf x}_2\|$, taking different values of $k$, considering $\delta = \sqrt{\frac{2}{\pi}}$. }
%\label{fig:bound}
%\end{figure}

The above relations are all statements of statistical expectation. In real applications, however, we must deal with a single realization of this kind of random process. To proceed, we note that the random variables $\left\{ \, d_{Lee}(Q_{x,A,U}({\bf x}_2)_i \, , Q_{x,A,U}({\bf x}_2)_i) \right\}_i$ are i.i.d. 
%In fact, using the Law of Large Numbers we know that, if $M \rightarrow \infty$, then
%\begin{multline*}
% \frac{\sum_{i=1}^M d_{Lee}\left( Q_{k,A,U}({\bf x}_1)_i \, , Q_{k,A,U}({\bf x}_2)_i \right)}{M} \\
%  \rightarrow \mathbb{E} \left[ d_{Lee}\left( Q_{k,A,U}({\bf x}_1)_i \, , Q_{k,A,U}({\bf x}_2)_i \right) \right]
%\end{multline*}

We define the \textit{Mean Lee Distance} as the empirical average of the Lee distances between components of $Q_{k,A,U}({\bf x}_1)$ and $Q_{k,A,U}({\bf x}_2)$:
\begin{multline*}
%\overline{d_{Lee}\left( Q_{k,A,U}({\bf x}_1) \, , Q_{k,A,U}({\bf x}_2) \right)} =
\overline{d_{Lee}\left( Q_{k,A,U}({\bf x}_1) \, , Q_{k,A,U}({\bf x}_2) \right)} =\\
 \frac{1}{M}\sum_{i=1}^M d_{Lee}\left( Q_{k,A,U}({\bf x}_1)_i \, , Q_{k,A,U}({\bf x}_2)_i \right)
\end{multline*}
To determine the value of $M$ such that the Mean Lee Distance between the hashes of ${\bf x}_1$ and ${\bf x}_2$ is a reasonable estimate of the {\em expected} value of the Lee distance, we can use the Hoeffding inequality and note that the Lee distance 
%between two elements in $\mathbb{Z}_k$ 
is a number between 0 and $\frac{k}{2}$ to prove that

\vspace{0.1cm}
\noindent \textbf{Proposition 3.} For $\delta = \sqrt{\frac{2}{\pi}}$, if $M \geq \frac{\log(2)\cdot (\beta +1) \cdot k^2}{8 \epsilon^2}$, then
\vspace{-0.3cm}
\begin{multline*}
\mathbb{P} \Bigg( \Big| \overline{d_{Lee}\left( Q_{k,A,U}({\bf x}_1) \, , Q_{k,A,U}({\bf x}_2) \right)}  \, - \\
 \mathbb{E} \left[ d_{Lee}\left( Q_{k,A,U}({\bf x}_1)_i \, , Q_{k,A,U}({\bf x}_2) \right)_i \right] \Big|  < \epsilon \Bigg) 
\geq 1 - \frac{1}{2^\beta}
\end{multline*}

Thus, the Mean Lee Distance is similar to its expected value with high probability if $M$ satisfies the described condition. For instance, if $\beta = 10$, $\epsilon = 0.5$ and $k = 8$, then with $M \geq 244$ we can obtain an estimate of the expected Lee Distance with precision 0.5 with probability more than 0.999.

\begin{figure}[t]
\centering
\includegraphics[trim={3.5cm 7.5cm 2cm 8cm},clip, scale=0.4]{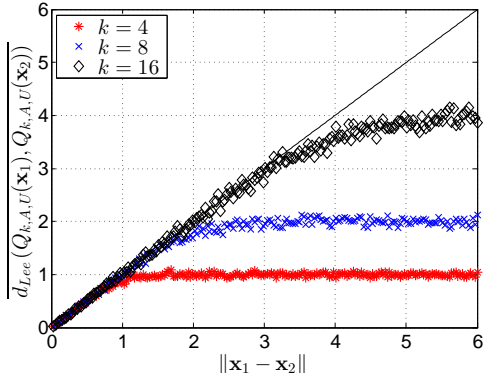}
\caption{Mean Lee distance between hashes as a function of $\|{\bf x}_1 - {\bf x}_2\|$ for $\delta = \sqrt{\frac{2}{\pi}}$, $N = 5000$ and $M = 500$.}
\label{fig:SimulatedCurve}
\vspace{-0.5cm}
\end{figure}

An important consequence of this result is that the value of $M$ required to obtain a good estimate of the expectation of the Mean Lee Distance between the hashes of ${\bf x}_1$ and ${\bf x}_2$, which in turn relates to the Euclidean distance between the signals, is not dependent on $N$, the dimensionality of the signal! Thus, we may even consider this kind of transform as a dimensionality-reduction technique or as a dimensionality-hiding hash.
Figure \ref{fig:SimulatedCurve} shows simulation plots of how the Mean Lee Distance approximates the euclidean distance, when the latter is lower than some threshold, for different values of $k$. This is absolutely consistent with the previous theoretical results. 

%In the following sections we describe some protocols to estimate the euclidean distance between two vectors using this technique.

\section{Protocol to Compute Distances}
We now describe a three-party protocol which uses SMH  to estimate the distance between two private signals without revealing them.

\vspace{0.2cm}
\noindent \textbf{Input}: Alice and Bob have ${\bf x}_1$ and ${\bf x}_2 \in \mathbb{R}^N$ respectively.

\noindent \textbf{Output}: Alice and/or Bob obtain an estimation of $\|{\bf x}_1 - {\bf x}_2\|$, provided it is lower than some threshold.

\noindent \textbf{Protocol}:

%\vspace{-0.1cm}
\begin{enumerate}[leftmargin=10pt]
\item Alice and Bob agree on a threshold $T$ and precision $\epsilon$, and compute  $k$ and $M$ using results from propositions 3 and 4.
\item Alice generates $(k,A,U)$ and sends them to Bob securely, this is $$a_{ij} \sim \mathcal{N}\left( 0, \frac{\pi}{2} \right) \quad \text{and} \quad u_i \sim \text{Unif}(0,k)$$
\item Alice computes $Q_{k,A,U}({\bf x}_1)$ and sends it to Charlie.
\item Bob computes $Q_{k,A,U}({\bf x}_2)$  and sends it to Charles.
\item Charles computes $d=\overline{d_{Lee}\left( Q_{k,A,U}({\bf x}_1) \, , Q_{k,A,U}({\bf x}_2) \right)}$ and sends $d$ to Alice and Bob.
\end{enumerate}

First, we can see that after this protocol no party gets anything more than the estimate of $\|{\bf x}_1 - {\bf x}_2\|$. Alice and Bob never see each others' vectors. Charlie never sees the plain vectors and receives just two vectors in $\mathbb{Z}_k^M$, where each one can be seen as a sequence of independent realizations of draws from a uniform distribution in $\mathbb{Z}_k$. Since Charles does not know $(A,U)$, he does not have any mechanism to reconstruct the actual vectors, and even extract any kind of information more than the distance between them.

Note that it is important that Charlie must not know $(A,U)$; otherwise he may reconstruct the original vectors, particularly if he has some knowledge of the domain of the  signals. 
%A simple attack is to apply the hash using the parameters on some known data and consider the reconstruction as the closer vector using the Mean Lee Distance. 
For the same reason Alice and Bob cannot directly share their hashes after sharing the key $(k,A,U)$. Provided these conditions are followed, the scheme can be shown to be information theoretically secure.

One drawback of this protocol is related to the key transmission. Indeed, if the vectors have a high dimension, transmitting the matrix $A$ may cause a communication overhead. However, we can make $A$ public because the security property of the hash reside on the random vector $U$. Hence, we can adapt the protocol as follows:
%publishing $A$, but Alice and Bob share $U$ and a permutation $\mathcal{P}$ of $M$ components. Then Alice and Bob compute $\mathcal{P}(Q_{k,A,U}({\bf x}_1))$  and $\mathcal{P}(Q_{k,A,U}({\bf x}_2))$ respectively and send them to Charles. Finally, Charles compute the Mean Lee Distance between the two received vectors.

\vspace{0.2cm}
\noindent \textbf{Input}: Alice and Bob have ${\bf x}_1$ and ${\bf x}_2 \in \mathbb{R}^N$ respectively. The $N\times M$ matrix $A$ is public.

\noindent \textbf{Output}: Alice and/or Bob obtain an estimation of $\|{\bf x}_1 - {\bf x}_2\|$, provided it is lower than some threshold.

\noindent \textbf{Protocol}:

\vspace{-0.2cm}
\begin{enumerate}[leftmargin=10pt]
\item Alice generates a $M$-dimensional vector $U$ with  $u_i \sim \text{Unif}(0,k)$, and a random permutation of $M$ components $\mathcal{P}$. Then Alice sends $(U,\mathcal{P})$ to Bob.
\item Alice computes $\mathcal{P}(Q_{k,A,U}({\bf x}_1))$ privately and sends it to Charles.
\item Bob computes $\mathcal{P}(Q_{k,A,U}({\bf x}_2))$ privately and sends it to Charles.
\item Charles computes $d=\overline{d_{Lee}\left( Q_{k,A,U}({\bf x}_1) \, , Q_{k,A,U}({\bf x}_2) \right)}$ and sends $d$ to Alice and Bob.
\end{enumerate}

Note that, despite the fact that vector $U$ results in hashes with uniform distribution, we added the permutation $\mathcal{P}$ in order to obstruct an exhaustive reconstruction by exploring all possible values of $U$. Following a similar analysis to the first protocol we can conclude that this protocol is secure.

However, these protocols have the disadvantage that the third party has access to the estimate of the euclidean distance between ${\bf x}_1$ and ${\bf x}_2$. Ideally we would like not to trust the third party to know this. In the next section we explore some solutions for this setting.

\section{Protocol to Hide Distance Value}

In order to not reveal the distance to the Third Party, we propose two protocols. The first does not consider the Third Party, and the distance computation between two real valued vectors is based on a Secure Hamming Distance computation between binary vectors.

The second protocol considers the idea of adding noise to the hashes in order to hide the true distance value.

\subsection{No Third Party}

One way to prevent the use of a Third Party is to establish a cryptographic protocol which lets us compute the Lee Distance privately. However, the computation of the Lee Distance involves  a minimum between two numbers, a task that can increase the computational complexity. 

Nevertheless, it is possible to reduce the Lee distance computation to a Hamming Distance computation. In fact, we can encode any element in $\mathbb{Z}_k$ as a vector in $\{0,1\}^{\frac{k}{2}}$. In particular, if $a\in \mathbb{Z}_k$, we define $c(a)\in \{0,1\}^{\frac{k}{2}}$ as follows:\\
If $a \leq \frac{k}{2}$, 
$$c(a)_i = \left\{ \begin{array}{ll} 1 & \text{ if } i\leq a\\ 0 & \text{ otherwise }\end{array} \right.$$
If $a > \frac{k}{2}$, 
$$c(a)_i = \left\{ \begin{array}{ll} 0 & \text{ if } i\leq a - \frac{k}{2} \\ 1 & \text{ otherwise }\end{array} \right.$$

For example, in $\mathbb{Z}_6$ the coding left
$$c(0) = \left[ \begin{array}{c}0\\0\\0 \end{array} \right] \quad 
c(1) = \left[ \begin{array}{c}1\\0\\0 \end{array} \right] \quad 
c(2) = \left[ \begin{array}{c}1\\1\\0 \end{array} \right] $$
$$c(3) = \left[ \begin{array}{c}1\\1\\1 \end{array} \right] \quad
c(4) = \left[ \begin{array}{c}0\\1\\1 \end{array} \right] \quad 
c(5) = \left[ \begin{array}{c}0\\0\\1 \end{array} \right] $$

This kind of coding has the property that
$$d_{Lee}(a,b) = d_{Hamming}(c(a)\, , c(b)) = \sum_{i=1}^{k/2} c(a)_i \oplus c(b)_i$$

Therefore, an element ${\bf z} \in \mathbb{Z}_k^M$ can be coded as $c({\bf z}) \in \{0,1\}^{M \cdot \frac{k}{2}}$, where
$$c({\bf z})^\top = \left[ \begin{array}{cccc} c(z_1)^\top,  &  c(z_2)^\top,  &  \cdots ,& c(z_M)^\top \end{array} \right]$$

Then, the Mean Lee Distance between ${\bf z}_1$ and ${\bf z}_2 \in \mathbb{Z}_k^M$ is equal to
$$\frac{d_{Hamming}(c({\bf z}_1), \, c({\bf z}_2 ) )}{M}$$

With this result we enable the estimation of the euclidean distance between two points in $\mathbb{R}^N$ using the Hamming distance of two binary vectors. As a consequence, we can replace the Third Party computation by any protocol which computes securely the Hamming distance between binary vectors; for example, in \cite{BinaryDistance:OT} defines a two party protocol based on Oblivious Transfer.

To summarize, we define the following protocol,

\vspace{0.1cm}
\noindent \textbf{Protocol:}

\vspace{-0.1cm}
\begin{enumerate}[leftmargin=10pt]
\item Alice generates $(k,A,U)$ and sends them to Bob.
\item Alice computes $c(Q_{k,A,U}({\bf x}_1))$ privately.
\item Bob computes $c(Q_{k,A,U}({\bf x}_2))	$ privately.
\item Alice and Bob apply a secure two party protocol to compute the Hamming distance $d$ between  $c(Q_{k,A,U}({\bf x}_1))$ and  $c(Q_{k,A,U}({\bf x}_2))$.
\item Alice and Bob compute the estimate of $\|{\bf x}_1 - {\bf x}_1 \|$ as $\frac{d}{M}$.
\end{enumerate}

Unlike most protocols based on homomorphic encryption to compute the euclidean distance, the complexity of the presented protocols does not depend on the dimensionality of the vector at the moment of applying the cryptographic technique. Hence, our proposal  for computing the euclidean distance between two vectors is to embed them in binary vectors and compute the Hamming distance between them. 

One problem of this protocol is the fact we lose information-theoretical privacy, unlike the previous ones.

\subsection{Obfuscating Information to the Third Party}

One alternative is to retain the Third Party but hide information from it through obfuscation. We simply propose adding noise to the corresponding hashes. The protocol is as follows:

\noindent \textbf{Protocol:}

\vspace{-0.2cm}
\begin{enumerate}[leftmargin=10pt]
\item Alice generates $(k,A,U)$, two independent vectors ${\bf z}_1$ and ${\bf z}_2$ in $\mathbb{Z}_k^P$, where each component is independent and distributed uniformly in $\mathbb{Z}_k$, and a permutation $\mathcal{P}$ of $M+P$ elements,  and sends $(k,A,U,{\bf z}_1,{\bf z}_2,\mathcal{P})$ to Bob.
\item Alice computes $\mathcal{P}\left( \left[ \begin{array}{c} Q_{k,A,U}({\bf x}_1)) \\ {\bf z}_1 \end{array} \right] \right)$ privately, and sends it to Charlie.
\item Bob computes $\mathcal{P}\left( \left[ \begin{array}{c} Q_{k,A,U}({\bf x}_2)) \\ {\bf z}_2 \end{array} \right] \right)$ privately, and sends it to Charlie.
\item Charlie computes the Mean Lee distance $d$ between the received vectors and sends it to Alice and Bob.

\item Alice and Bob compute privately the Mean Lee distance $\tilde{d}$ between ${\bf z}_1$ and ${\bf z}_2$ and compute $$\frac{(M+P)\cdot d - P\cdot \tilde{d}}{M}$$ obtaining the estimate of $\|{\bf x}_1 - {\bf x}_2\|$.
\end{enumerate}

\noindent It is easy to see that Mean Lee distance between $Q_{k,A,U}({\bf x}_1))$ and $Q_{k,A,U}({\bf x}_1))$ is equal to the last expression of the protocol. One of the advantages of this protocol is the fact that, when $P$ is large enough, $\mathcal{P}\left( \left[ \begin{array}{c} Q_{k,A,U}({\bf x}_2)) \\ {\bf z}_2 \end{array} \right] \right)$ is indistinguishable from a vector with uniformly distributed i.i.d components, therefore, the value of $d$ should be close to $\frac{k}{4}$ for every pair ${\bf x}_1$ and ${\bf x}_2$. 

A natural drawback is that the number of parameter to run the protocol, which may cause a computational overhead.

\section{Conclusions}
In this paper we have presented a random transformation, that given a threshold $T$, can generate hashes from vectors that preserving the euclidean distance between them if it is smaller than $T$. These hashes are uninformative if the random parameters of the function are unknown. 

With this kind of transformation, we are able to describe protocols to compute distances privately and efficiently. In fact, the proposed transformation is not only uninformative, it can be see as a transformation which reduces the data dimension and preserves the euclidean distance through the Lee distance output space. 

Although the fact that the distance is preserved until some threshold seems to be a drawback, if both Alice and Bob know the maximum possible distance between their vectors, then they can set an appropriate value of $k$ which lets preserve distances to any desired threshold. We also describe how to hide the distance value from  the third party, action that may increase the computational cost of the protocol, as it is expected.
 
We still have many questions. How can we extend these protocols to multiparty versions? How easy is a reconstruction of ${\bf x}$ knowing $Q_{k,A,U}({\bf x})$ and $(k,A,U)$? We have also trusted the third party to not collude with Alice or Bob. Can this restriction be lifted?

\section{Appendix - Proofs}

\vspace{0.1cm}
\noindent \textbf{Theorem 2.} $\forall {\bf x}_1 \, , {\bf x}_2 \in \mathbb{R}^N$, $\forall i \in \{1,...,M\}$,  $\forall k$ even, then we obtain the following relation between the $i^{\rm th}$ components of $Q_{k,A,U}({\bf x}_1)$ and $Q_{k,A,U}({\bf x}_2)$:
\begin{multline*}
 \mathbb{E} \left[ d_{Lee}\left(Q_{k,A,U}({\bf x}_1)_i,Q_{k,A,U}({\bf x}_2)_i \right) \right] = \\
 \frac{k}{4} -  \frac{2k}{\pi^2} \sum\limits_{j=1}^\infty \frac{1}{(2j-1)^2} e^{-2 \left( \frac{\pi \|{\bf x}_1-{\bf x}_2\|(2j-1)}{\delta k}\right)^2}  
\end{multline*}

\vspace{0.1cm}

\noindent \textbf{Proof}

\begin{enumerate}
\item[]
We can consider, with loss of generality, that $M=1$.

We have that 

$$\mathbb{P}\left(d_{Lee}(Q_{k,A,U}({\bf x}_1), Q_{k,A,U}({\bf x}_2))\leq j \, \Bigg| \, \left\| A({\bf x}_1-{\bf x}_2) \right\| \right)$$ 

is given by the function $g_k$ depending on $L=\left\| A({\bf x}_1-{\bf x}_2) \right\|$, described in the following figure

\includegraphics[scale=0.45]{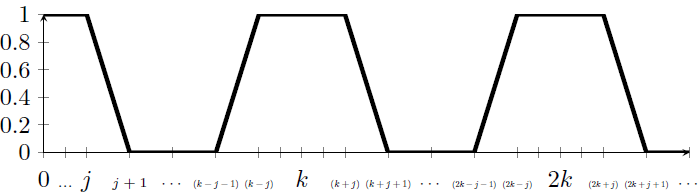}

Besides, we know that the density function of $L$ is
$$f_L(l) = \sqrt{\frac{2}{\pi}} \frac{\delta}{\|{\bf x}_1-{\bf x}_2\|} e^{-\frac{\delta^2l^2}{2\|{\bf x}_1-{\bf x}_2\|^2}}$$

Then
$$\mathbb{P}\left(d_{Lee}(Q_{k,A,U}({\bf x}_1),Q_k({\bf x}_2))\leq j \right) = \int_0^\infty g_k(l) f_L(l) dl$$

But, since both $g_k$ and $f_L$ are positive, we can extend these functions over real line, defining $\tilde{g}_k$ as
$$ \tilde{g}_k(x) = \left\{ \begin{array}{lll} g_k(x) & \mbox{ if } & x\geq 0 \\ g_k(-x) & \mbox{ if } & x<0 \end{array} \right. $$

Similarly we define $\tilde{f}_L$. Then
$$\mathbb{P}\left(d_{Lee}(Q_{k,A,U}({\bf x}_1),Q_{k,A,U}({\bf x}_2))\leq j \right) = \frac{1}{2}\int_{-\infty}^\infty \tilde{g}_k(l) \tilde{f}_L(l) dl$$

However, we can see that
$$ \tilde{g}_k(l) = \textrm{train}_k \ast h \, (l)$$

where $$\textrm{train}_k(l) = \sum_{i=-\infty}^\infty \delta(l-ik)$$ 

and $$h(x) = {\scriptsize \left\{ \begin{array}{cl} x+j+1& \mbox{ if } x\in [-j-1,-j]\\ &  \\ 1 & \mbox{ if } x\in [-j,j]\\ & \\ -x+j+1 & \mbox{ if } x \in [j,j+1] \\ & \\ 0 & \mbox{ otherwise} \end{array} \right. }$$

Additionally, using the Parseval's theorem we have
$$\mathbb{P}\left(d_{Lee}(Q_{k,A,U}({\bf x}_1),Q_{k,A,U}({\bf x}_2))\leq j \right) = \frac{1}{2}\int_{-\infty}^\infty \hat{\tilde{g}}_k(\xi) \hat{\tilde{f}}_L(\xi) d\xi$$

where $\hat{\tilde{g}}_k$ and $\hat{\tilde{f}}_L$ are the Fourier transform of $\tilde{g}_k$ and $\tilde{f}_L$ respectively.
But, using the definition of $\tilde{g}_k$ we have

\begin{eqnarray*}
\hat{\tilde{g}}_k(\xi) & = & \hat{h}(\xi) \cdot \hat{\textrm{train}}_k (\xi) \\ 
& = & \frac{1}{2\pi^2 \xi^2} \left( \cos (2 \pi \xi j) - \cos (2 \pi \xi (j+1)) \right)\\
& & \qquad \cdot \left( \frac{1}{k} \sum_{i=-\infty}^\infty \delta\left( \xi - \frac{i}{k} \right) \right) 
\end{eqnarray*}

because
$$\hat{h}(\xi) = \frac{1}{2\pi^2 \xi^2} \left( \cos (2 \pi \xi j) - \cos (2 \pi \xi (j+1)) \right)$$

On the other hand, since $\tilde{f}_L$ is a gaussian, it is easy to see 
$$ \hat{\tilde{f}}_L(\xi) = 2 e^{-2 \left( \frac{\pi \|{\bf x}_1-{\bf x}_2\| \xi}{\delta} \right)^2} $$

Therefore
\begin{align*}
& \mathbb{P}\left(d_{Lee}(Q_{k,A,U}({\bf x}_1),Q_{k,A,U}({\bf x}_2))\leq j \right) \\
& =  \int_{-\infty}^\infty  \frac{1}{k} \frac{1}{2\pi^2 \xi^2} \left( \cos (2 \pi \xi j) - \cos (2 \pi \xi (j+1)) \right) \\
& \qquad  \qquad \qquad \cdot \sum_{i=-\infty}^\infty \delta \left( \xi - \frac{i}{k}\right) e^{-2\left( \frac{\pi \|{\bf x}_1-{\bf x}_2\| \xi}{\delta} \right)^2} d\xi \\
& =  \frac{1}{k} \sum\limits_{i=-\infty}^\infty \frac{1}{2 \left( \frac{\pi i}{k}\right)^2} \left[ \cos \left( 2 \pi j \frac{i}{k} \right) - \cos \left( 2 \pi (j+1) \frac{i}{k} \right) \right] \\
& \qquad \qquad \qquad \cdot e^{-2\left( \frac{\pi \|{\bf x}_1-{\bf x}_2\| i}{\delta k} \right)^2}\\
& = \frac{k}{\pi^2} \sum\limits_{i=1}^\infty \frac{\left[ \cos \left( 2 \pi j \frac{i}{k} \right) - \cos \left( 2 \pi (j+1) \frac{i}{k} \right) \right]}{i^2}  e^{-2\left( \frac{\pi \|{\bf x}_1-{\bf x}_2\| i}{\delta k} \right)^2} \\
& \qquad \qquad + \frac{2j+1}{k}
\end{align*}

Finally, noting that if $k$ is even, we have 
$$0 \leq d_{Lee}(Q_{k,A,U}({\bf x}_1),Q_{k,A,U}({\bf x}_2)) \leq \frac{k}{2}$$
And
\begin{align*}&\mathbb{E} \left[ d_{Lee}(Q_{k,A,U}({\bf x}_1),Q_{k,A,U}({\bf x}_2)) \right] \\
& = \sum_{i=0}^{k/2} 1- \mathbb{P}\left(d_{Lee}(Q_{k,A,U}({\bf x}_1),Q_{k,A,U}({\bf x}_2))\leq j \right) 
\end{align*}
we can prove that
\begin{multline*}
 \mathbb{E} \left[ d_{Lee}(Q_{k,A,U}({\bf x}_1),Q_{k,A,U}({\bf x}_2)) \right] = \\
 \frac{k}{4} -  \frac{2k}{\pi^2} \sum\limits_{i=1}^\infty \frac{1}{(2i-1)^2} e^{-2 \left( \frac{\pi \|{\bf x}_1-{\bf x}_2\|(2i-1)}{\delta k}\right)^2}  
\end{multline*}

\hfill $\square$
\end{enumerate}

\noindent \textbf{Corollary}
\begin{multline*}
\frac{k}{4}-\frac{k}{4}e^{-2 \left( \frac{\pi \|{\bf x}_1-{\bf x}_2\|}{\delta k}\right)^2} \leq \\
\mathbb{E}[d_{Lee}(Q_{k,A,U}({\bf x}_1),Q_{k,A,U}({\bf x}_2))] \leq \\ \frac{k}{4}-\frac{2k}{\pi^2}e^{-2 \left( \frac{\pi \|{\bf x}_1-{\bf x}_2\|}{\delta k}\right)^2}
\end{multline*}

\noindent \textbf{Proof}

\begin{enumerate}
\item[]

Considering just the first term of the series, we have
$$\mathbb{E}[d_{Lee}(Q_{k,A,U}({\bf x}_1),Q_{k,A,U}({\bf x}_2))] \leq  \frac{k}{4}-\frac{2k}{\pi^2}e^{-2 \left( \frac{\pi \|{\bf x}_1-{\bf x}_2\|}{\delta k}\right)^2}$$

On the other hand, we can notice that
\begin{align*}
&  \sum_{i=1}^\infty \frac{1}{(2i-1)^2} e^{-2\left(\frac{\pi \|{\bf x}_1-{\bf x}_2\|(2i-1)}{\delta k} \right)^2}\\
& \qquad \leq  \sum_{i=1}^\infty\frac{1}{(2i-1)^2}  e^{-2\left(\frac{\pi \|{\bf x}_1-{\bf x}_2\|}{\delta k} \right)^2} \\
& \qquad =  e^{-2\left(\frac{\pi \|{\bf x}_1-{\bf x}_2\|}{\delta k} \right)^2} \sum_{i=1}^\infty \frac{1}{(2i-1)^2} \\
& \qquad =  e^{-2\left(\frac{\pi \|{\bf x}_1-{\bf x}_2\|}{\delta k} \right)^2} \frac{\pi^2}{8} 
\end{align*}
Then, applying this inequality, the second inequality holds.

\hfill $\square$
\end{enumerate}

%-------------------------------------------------------------------------------------
%  Proposition 1
%-------------------------------------------------------------------------------------
\vspace{2cm}
\noindent \textbf{Proposition 1.}
Let $\delta = \sqrt{\frac{2}{\pi}}$. Defining the error 
\begin{multline*}
 \varepsilon(\|{\bf x}_1 - {\bf x}_2 \|,k) : = \\
 \left| \mathbb{E}[d_{Lee}\left(Q_{k,A,U}({\bf x}_1)_i,Q_{k,A,U}({\bf x}_2)_i \right)] -  \| {\bf x}_1 - {\bf x}_2 \| \right| 
\end{multline*}
the following relation holds:
\begin{equation*}
\varepsilon(\|{\bf x}_1 - {\bf x}_2 \|, k) \leq F(\|{\bf x}_1 - {\bf x}_2 \|, k) 
\end{equation*}
where
$$F(t,k) = t \cdot \exp \left( - \frac{k^2}{4 \pi t^2} \right)$$

\noindent \textbf{Proof}

\begin{enumerate}
\item[]Using the same scheme as in the proof of theorem 2, we can see that 
$$ \mathbb{E}\left[ d_{Lee} \left( Q_{k,A,U}({\bf x}_1)_i, \, Q_{k,A,U}({\bf x}_2)_i \right) \right] = \int_0^\infty f_L(u) w(u) du$$

where $f_L$ is given by
$$f_L(u) = \sqrt{\frac{2}{\pi}} \frac{\delta}{\|{\bf x}_1-{\bf x}_2\|} e^{-\frac{\delta^2u^2}{2\|{\bf x}_1-{\bf x}_2\|^2}}$$
and $w$ by the function shown in the following plot

\includegraphics[scale=0.45]{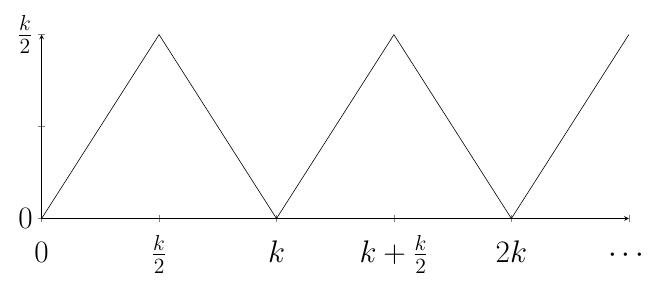}

It is easy to see that 
$$w(u) \leq u \qquad \forall u$$

then
\begin{eqnarray*}
\mathbb{E}\left[ d_{Lee} \left( Q_{k,A,U}({\bf x}_1)_i, \, Q_{k,A,U}({\bf x}_2)_i \right) \right] & \leq & \int_0^\infty f_L(u) u\, du\\
 & = & \sqrt{\frac{2}{\pi}} \cdot \frac{\|{\bf x}_1 - {\bf x}_2\|}{\delta}
\end{eqnarray*}

Besides, noting that

$$  d_{Lee} \left( Q_{k,A,U}({\bf x}_1)_i, \, Q_{k,A,U}({\bf x}_2)_i \right)  \geq 0$$

we can write
\begin{multline*}
\left| \mathbb{E}\left[ d_{Lee} \left( Q_{k,A,U}({\bf x}_1)_i, \, Q_{k,A,U}({\bf x}_2)_i \right) \right] -  \sqrt{\frac{2}{\pi}} \cdot \frac{\|{\bf x}_1 - {\bf x}_2\|}{\delta} \right| = \\
\left| \int_0^\infty f_L(u) \cdot ( w(u)-u) \,du\right|
\end{multline*}

but

$$\left| \int_0^\infty f_L(u) \cdot ( w(u)-u) du\right| \leq \int_{\frac{k}{2}}^\infty f_L(u)\cdot u \,du$$

and we have

$$ \int_{\frac{k}{2}}^\infty f_L(u)\cdot u \,du = \sqrt{\frac{2}{\pi}} \frac{\|{\bf x}_1 - {\bf x}_2\|}{\delta} e^{- \frac{\delta^2 k^2}{8 \|{\bf x}_1 - {\bf x}_2 \|^2}}$$

Hence, setting $\delta =  \sqrt{\frac{2}{\pi}}$ we have

\begin{multline*}
\Bigg| \mathbb{E}\left[ d_{Lee} \left( Q_{k,A,U}({\bf x}_1)_i, \, Q_{k,A,U}({\bf x}_2)_i \right) \right] -  \|{\bf x}_1 - {\bf x}_2\| \Bigg| \leq \\
\|{\bf x}_1 - {\bf x}_2\| \cdot e^{- \frac{k^2}{4 \pi \|{\bf x}_1 - {\bf x}_2 \|^2}}$$
\end{multline*}

And defining

$$F(\|{\bf x}_1 - {\bf x}_2 \| = \|{\bf x}_1 - {\bf x}_2\| \cdot e^{- \frac{k^2}{4 \pi \|{\bf x}_1 - {\bf x}_2 \|^2}}$$

we have the result.

\hfill $\square$

\end{enumerate}

%-------------------------------------------------------------------------------------
%  Proposition 2
%-------------------------------------------------------------------------------------

\vspace{2cm}
\noindent \textbf{Proposition 2.} $\forall \epsilon >0$ , $\forall T>0$, $\exists k$ even,  $\forall \|{\bf x}_1 - {\bf x}_2\| <T$
$$ \varepsilon(\|{\bf x}_1 - {\bf x}_2\|,k) < \epsilon$$

\noindent \textbf{Proof}

\begin{enumerate}
\item[] To see this is enough to verify that $F(t,k)$ is increasing in $t$ and when $k$ tends to infinity $F(t,k)$ tends to 0.

To see that $F$ increases in $t$ we can just compute the derivative and it is easy to verify that is positive.

On the other hand, it is clear that, when $k$ tends to infinity, $F(t,k)$ tends to zeros for every value of $t$

\hfill $\square$
\end{enumerate}

%-------------------------------------------------------------------------------------
%  Proposition 3
%-------------------------------------------------------------------------------------

\vspace{1cm}
\noindent \textbf{Proposition 3.} For $\delta = \sqrt{\frac{2}{\pi}}$, if $M \geq \frac{\log(2)\cdot (\beta +1) \cdot k^2}{8 \epsilon^2}$, then
\vspace{-0.3cm}
\begin{multline*}
\mathbb{P} \Bigg( \Big| \overline{d_{Lee}\left( Q_{k,A,U}({\bf x}_1) \, , Q_{k,A,U}({\bf x}_2) \right)}  \, - \\
 \mathbb{E} \left[ d_{Lee}\left( Q_{k,A,U}({\bf x}_1)_i \, , Q_{k,A,U}({\bf x}_2) \right)_i \right] \Big|  < \epsilon \Bigg) 
\geq 1 - \frac{1}{2^\beta}
\end{multline*}

\noindent \textbf{Proof}

\begin{enumerate}
\item[] The Hoeffding inequality establish that, if $X_i \in [a_i, b_i]$, and $X=\sum_i X_i$, then
 $$\mathbb{P}\left( |\mathbb{E}(X)-X|\geq \delta \right) \leq 2 \exp \left( \frac{-2 \delta^2}{\sum_i (b_i-a_i)^2} \right)$$
In our case, noting that 
$$0 \leq d_{Lee}(Q_{k,A,U}({\bf x}_1), Q_k({\bf x}_2)) \leq \frac{k}{2}$$ we have

 \begin{multline*} \mathbb{P}\Bigg( \Bigg|\mathbb{E}(d_{Lee}(Q_{k,A,U}({\bf x}_1)_i, Q_{k,A,U}({\bf x}_2)))\\
 \quad -\frac{\sum_{i=1}^M d_{Lee}\left( Q_{k,A,U}({\bf x}_1)_i , Q_{k,A,U}({\bf x}_2)_i \right)}{M}\Bigg|\geq \epsilon  \Bigg)\\
 \leq \quad 2 \exp \left( \frac{-8 M \epsilon^2}{ k^2} \right)
  \end{multline*} 
  
Then, setting 
 
$$  2 \exp \left( \frac{-8 M \epsilon^2}{ k^2} \right) = 2^{-\beta}$$

We can conclude that a sufficient condition is to consider 

$$M \geq \frac{\log(2) \cdot(\beta+1) k^2}{8 \epsilon^2}$$

\hfill $\square$
\end{enumerate}


\begin{thebibliography}{1}

\bibitem{manaspathak:smc}
M. Pathak and B. Raj, ``Privacy-Preserving Speaker Verification and Identification Using Gaussian Mixture Models.'' IEEE Transactions on Audio, Speech and Language Processing, Vol 21:2,  pp. 397-406, 2013.

\bibitem{fhm:latest}
M. Naehrig, K. Lauter and V. Vaikuntanathan, ``Can homomorphic encryption be practical?.'' Proceedings of the 3rd ACM workshop on Cloud computing security workshop, 2011.

\bibitem{RANE:LSHinsecure}
S. Rane and P. T. Boufounos, ``Privacy-Preserving Nearest Neighbor Methods: Comparing Signals Without Revealing Them.'' IEEE Signal Processing Magazine, Vol. 30(2), pp. 18-28, 2013.


\bibitem{PolyDistance}
Y. Wang, P. Ishwar and S. Rane, ``Information-theoretically secure three-party computation with one active adversary.'' 2012. [Online]. Available: http://arxiv.org/abs/1206.2669


\bibitem{LSH:Indyk}
P. Indyk and R. Motwani. ``Approximate Nearest Neighbors: Towards Removing the Curse of Dimensionality.'' Proceedings of 30th Symposium on Theory of Computing, 1998.

%\bibitem{LSH:Andoni}
%A. Andoni and P. Indyk, ``Near-Optimal Hashing Algorithms for Approximate Nearest Neighbor in High Dimensions.'' Communications of the ACM, Vol. 51(1), pp. 117–122, 2008.

\bibitem{LSHfordistance}
M. Datar, N. Immorlica, P. Indyk and V.S. Mirrokni, ``Locality-Sensitive Hashing Scheme Based on p-Stable Distributions.'' Proceedings of the Symposium on Computational Geometry, 2004.


\bibitem{Jimenez:wifs}
A. Jimenez, B. Raj, J. Portelo and I. Trancoso, ``Secure Modular Hashing.'' Proceedings of WIFS, 2015.

\bibitem{LeeDistance}
E. Deza and M. Deza, ``Dictionary of Distances'', Elsevier, 2014.

%\bibitem{Petros:SBE}
%P. T. Boufounos and S. Rane, ``Secure Binary Embeddings for Privacy Preserving Nearest Neighbors.'' Proceedings of WIFS, 2011.

\bibitem{BinaryDistance:OT}
J. Bringer, H. Chabanne and A. Patey, ``SHADE: Secure HAmming DistancE Computation from Oblivious Transfer.'' Financial Cryptography and Data Security
Volume 7862 of the series Lecture Notes in Computer Science pp 164-176. 2013.



\end{thebibliography}
\end{document}